# The effects of graphite nanoparticles, phase change material, and film cooling on the solar still performance

S.W. Sharshir [a,b,c,#], Guilong Peng [a,b,#], Lirong Wu [a,b], Nuo Yang [a,b,*], F.A. Essa [c], A.E. Kabeel [d]

[a] State Key Laboratory of Coal Combustion, Huazhong University of Science and Technology, Wuhan 430074, China

[b] Nano Interface Center for Energy (NICE), School of Energy and Power Engineering, Huazhong University of Science and Technology, Wuhan 430074, China

[c] Mechanical Engineering Department, Faculty of Engineering, Kafrelsheikh University, Kafrelsheikh, Egypt.

[d] Mechanical Power Engineering Department, Faculty of Engineering, Tanta University, Tanta, Egypt

#S.W. S. and G. P. contributed equally to this work.
*Corresponding authors: N.Y. (nuo@hust.edu.cn)

In this paper, we performed four modifications on the solar still, as (A) adding the graphite nanoparticles, (B) the graphite nanoparticles together with the phase change material (PCM), (C) the graphite nanoparticles together with glass film cooling, and (D) graphite nanoparticles with both PCM and glass film cooling. The effects of modifications are measured and compared with each other. The productivities of modified (A), (B), (C), and (D) solar stills are enhanced by about 50.28%, 65.00%, 56.15% and 73.80%, respectively, as compared with the conventional solar still. The influences of saline water depths on the performance of modifications (A) and (B) are also considered. Results revealed that the best output yield is obtained for 0.5cm water depth for all solar stills.

In the past few decades, clean water supplies have become a lot more critical due to excessive use and increasing contamination of natural water sources. Moreover, the demand for drinking water in the world is increasing and regulations on drinking water quality have become a lot more stringent[1]. By 2025, there will be a big problem in water vulnerability for more than half of the world population[2] Hence, people have to use efficient methods to produce freshwater. Solar still desalination is one of these methods. Solar still is a device having the advantages of easily fabricating, cheap, no specific skills to operate, approximately no maintenance and no need of conventional energy. On the other hand, it is not popularly used because of its low productivity.

Many works had been carried out to improve the productivity of the solar still such as plastic solar water purifier[3], regenerative solar desalination unit[4], asymmetric greenhouse type solar still with some mirrors[5], reflector with corrosion free absorbers[6], flat plate collectors[7, 8, 9], wick type still[10], triple-basin still[11], capillary film[12], multi effect solar still with thermal energy recycle[13], solar water collector[14, 15], black gravel or black rubber[16, 17], sponge cubes[18], single and double slope solar still[19], electrical blower[20], baffle suspended absorber[21], and energy storing and wick materials[22- 24].

Besides to the modifications we mentioned above, researchers also found that the condensation rate and productivity can be increased by increasing the water-glass temperature difference which can be kept up as a maximal value by increasing the cooling water flow rate and decreasing the inlet cooling water temperature[25- 29]. Meanwhile, phase change materials (PCMs), which are able to store and release energy, are universally used in solar systems. The heat is absorbed (liberated) during melting (solidifying). It stores the heat energy during the daytime and releases it during the night and cloudy days. Many researchers had used PCMs as an improving parameter of solar still productivity. The effect of using latent heat thermal energy storage system (LHTESS) through two cascade solar stills is investigated by Tabrizi et al.[30] Results obtained that the productivity of basin still with LHTESS is slightly lower than the still without LHTESS. The theoretical study of solar still with and without PCMs is carried out by Dashtban and Tabrizi[31]. The daily productivity reached 6.7 and 5.1 kg/m$^2$ with and without PCMs respectively. Ansari et al.[32] examined a passive solar still integrated with a PCMs beneath the basin liner.

Recently, with the development of nanotechnology, the nanofluid has attracted the attention of many researchers in solar desalination field. Nanofluid has a lot of special properties compared to its base liquid such as high thermal conductivity[33-40], high solar radiation absorptivity[41], which are helpful parameters to increase the productivity of the solar still. Some researchers studied the effects of using different types of nanofluid on the productivity of solar still. The efficiency of solar still was increased by 29% when using a violet dye as indicated by Nijmeh et al.[42]. Elango et al.[43] conducted an experimental study to improve the productivity of single basin single slope solar still using nanofluid. The productivity of basin still with aluminum oxide ($Al_2O_3$) nanofluid was improved by 29.95%, while the productivities of solar stills with tin oxide ($SnO_2$) and zinc oxide ($ZnO$) nanofluids were enhanced by 18.63% and 12.67% higher than that without nanofluid respectively.

Kabeel et al.[44, 45] investigated the effects of using aluminum oxide nanoparticles and providing vacuum with integrating an external condenser to the solar still under the Egyptian conditions[45]. Results obtained an increase of 53.2% in the total daily productivity when providing vacuum inside the basin still and an increase of 116% when using the aluminum oxide nanoparticles with providing vacuum. Sahota and Tiwari[46] conducted an experimental and theoretical study to improve the productivity of double slope solar still (DSSS) using $Al_2O_3$ nanoparticles. The productivity of DSSS with aluminum oxide ($Al_2O_3$) nanofluid was improved by 12.2% and 8.4% at 35 kg and 80 kg base fluid respectively, with 0.12% concentration of Al2O3 nanoparticles.

From the above literature review, it is observed that the effects of using either some new nanomaterials or coupling the nanomaterials with PCMs and film cooling are not investigated. In our experiment, we chose graphite micro-flakes as our experimental object in consideration of its relative high thermal conductivity[47], low cost, and low density as compared to most of nanomaterials. Therefore, the objectives of this work are to enhance the solar still performance by the modifications of: (A) the nanofluid of graphite nanoparticles; (B) the nanofluid combined with PCM as thermal storage materials; (C) the nanofluid combined with film cooling (glass cover cooling); and (D) the nanofluid combined with both PCMs and film cooling. In addition, we also studied the effects of water depths on the solar still performance with modifications (A) and (B).

**Results**

Depending on the weather conditions, the ambient temperature is varied from 20 to 28 °C and the wind speed is varied from 0.1 to 5 m/s while the solar intensity is varied from 30 to 880 W/m$^2$ at different days.

**Effect of using the nanofluid of graphite nanoparticles[modification (A)]**

The effect of the graphite nanofluid on the performance of solar still is shown in Figs. 3. The temperature and solar radiation curves have the same trend for all the day of tests as shown in Fig. 3a. It is obtained from the figures that the water temperatures and glass temperatures of the graphite nanofluid still are higher than those of the conventional still by 0 – 4ºC and 0 – 2 ºC, respectively this is because, nanofluid (water+ graphite) have higher thermal conductive and also absorbs more solar radiation than the water only. Hence, the evaporation and production rates are better in modified (A) still than that of conventional still . The variations of hourly freshwater productivity for modification (A) and conventional solar stills are presented in Fig. 3b. It is found from the figure that the amount of accumulated water of the modified (A) still is higher than that of the conventional still. The evaporation rate is increased due to increasing the heat transfer rate and water temperature because of the nanofluids addition. When using the graphite nanoparticles, the productivity of modified (A) still was enhanced by about 50.28 % as compared to the conventional still at a brine depth of 0.5 cm.

**Effect of using the graphite nanoparticles and PCM [modification (B)]**

The hourly temperature variations and solar radiation for the solar still with PCM and conventional still are illustrated in Fig. 3c. While, the hourly distillate variation for the modified (B) solar still and conventional still is shown in Fig. 3d. From Fig 3 it is observed that, the solar radiation, temperatures of saline water, glass cover, and phase change material PCM (Paraffin Wax) are measured and drawn to evaluate the performance of the desalination unit. It is seen from Fig. 3c that all temperatures of saline water, glass cover, and PCM are increased gradually with the increase of solar intensity and hence, the solar still hourly productivity is increased also with the increase of solar radiation.

It can be observed from Fig. 3d that the productivity trend is similar to that of the temperature trend as observed from Fig. 3c. It is also observed from the figures that the peaks of temperatures and productivity are late after the peak of solar radiation. This is because that some of heat inside the solar still are stored as a sensible and latent heat within the PCM and this needs longer time and larger amount of energy to rise the temperatures. Therefore, the temperatures and productivity of conventional still are higher from about 9:00 am to 12:00/1:00 pm as shown in Fig.3. After that time (13:00 PM), the modified (B) solar still has a higher water and glass cover temperatures than that of the conventional type. Furthermore, it is obtained from Fig. 3c that the water and glass cover temperatures of the conventional basin drop very fast after 2:00 pm, while the PCM inside the modified (B) solar still works as a heat source and hence the temperatures decrease slowly with time.

Results also indicated that the conventional basin still has a very low productivity during the nighttime (from sunset of a day to sunrise of the next day). On the other hand, for the modified (B) still, a considerable amount of fresh water productivity is continued to be produced during the nighttime as well as during low intensity solar radiation periods. In addition, the daily productivities recorded approximately 529 and 873 ml/day for conventional and modified (B) solar stills, respectively. So, the modified (B) solar still has higher productivity by 65% than that of the conventional type at a brine depth of 0.5 cm.

**Effect of using the graphite nanoparticles and film cooling [modification (C)]**

As mentioned before, using the graphite nanoparticles increases the evaporation rate due to the improved heat transfer and irradiation absorption characteristics of the resulted nanofluid. As a result, the glass temperature is increased due to the high latent heat of vaporization received. Hence, this causes low productivity because the temperature difference between water and glass is decreased. So, our target is to increase the water-glass temperature difference to increase the distillate productivity. So, we used a flowing cold water over the glass cover to catch some heat stored in the glass by conduction and convection. Consequently, the glass cover temperature is get back down to keep the water-glass temperature difference as large as possible. Hence, the condensation and productivity rates are increased. Another useful result of using the cooling film is the continuous self-cleaning of the glass cover. Therefore, the solar still efficiency is maintained with high levels.

Fig. 4a shows the variation of solar radiation, basin water, glass cover and ambient temperatures for modified (C) and conventional stills. The hourly distillate variations for the modified (C) and conventional stills are also shown in Fig. 4b. Experimentations were conducted using the flow rate of 0.03kg/s for the cold water flowing over the glass cover. From the figure, it can be seen that the glass cover temperature of modified (C) still is less than that of conventional still by about 1–26 °C due to cooling of modified (C) glass cover. Results indicate that the temperature difference between the glass temperature and brine temperature for modified conventional still increases with cooling (reaches about 27 °C) and without cooling (reaches about 11 °C). It is found that the productivity of the modified (C) solar still is increased approximately by 56.15 % as compared to the conventional still at a brine depth of 0.5 cm.

**Effect of using the graphite nanoparticles, PCM and film cooling [modification (D)]**

Using the PCM and graphite nanoparticles increases the evaporation rate and using the cooling film individually increases the condensation rate and hence they increase the productivity of the solar still. So, in this part, the authors investigated the performance of the solar still when using both of PCM and graphite nanoparticles with film cooling. The hourly variations of temperatures of water ($T_w$), glass cover ($T_g$), PCM ($T_{pcm}$) of Praveen Wax as a PCM and productivity are obtained in Fig. 4. It can be obtained from the Fig. 4c, that as the solar radiation increases with time, the PCM temperature increases because of the increased heat transfer by conduction from the black metal pipes to the PCM. The absorbed heat by the PCM makes it melting after 3 and 5 h from the exposure of solar still to solar radiation in the morning. After 14:00, PCM begins to discharge the heat stored and keeps the water to be warmer than that of the conventional still. This causes a significant difference in productivity between the modified (D) and conventional solar stills during the sunset as shown in Fig. 4d.

Also as mentioned above, using the cooling water film flowing over the glass cover makes the water-glass temperature difference larger than that without using the film cooling. Hence, the condensation and production rates are high. Experimentations also were conducted using the same flow rate in modification (C) of 0.03 kg/s for the cold water flowing over the glass cover. From the figure, it can be seen that the glass cover temperature of modified (D) is less than that of conventional solar still by about 1–22 °C due to cooling of modified (D) glass cover. Results indicate that the temperature difference between the glass temperature and brine temperature for modified (D) increases with cooling (reaches about 28 °C) and without cooling (reaches about 11 °C). It is also found that the productivity of the modified (D) is increased approximately by 73.8 % as compared to the conventional still at a brine depth of 0.5 cm. Comparison between present study and different research works about solar still with nanofluids are illustrated in Table 2.

**Effect of water depth on modifications (A) and (B)**

Two cases under investigations were done with varying the brackish water depth (0.5, 1, and 2 cm) to get the optimum water depth for maximum distilled water when compared to the conventional still.

Fig. 5 shows a comparison between the modified (A) and modified (B) with conventional solar still at different operating conditions. Different depths effects, for basin water, on the productivity were tested. The figure showed that the more decrease in basin water depth, the more increase in productivity for the two tested cases, and 0.5cm is the best depth. Also, it is observed that the percentage of increase in productivity for the solar still with PCM and using graphite nanoparticles is greater than that of solar still using graphite nanoparticles only regardless the depth.

**Discussion**

The performance of solar still with graphite nanoparticles, phase change material, and film cooling was experimentally investigated. Besides, the influence of saline water depths (0.5, 1, and 2 cm) was studied. The distillate productivity, as well as the thermal performance, of a conventional solar still can be improved through the design modifications. Based on the measurements, the detailed results are obtained as:

1- The productivity of solar still with graphite nanoparticles, modification (A), is 50.28 % higher than that of the conventional still.
2- The productivity of the solar still is improved by using graphite nanoparticles and PCM as energy storage materials, modification (B), with 65% higher than the productivity of the conventional still.
3- The productivity of solar still with graphite nanoparticles and glass cover cooling, modification (C), is 56.15% higher than that of the conventional still.

4- When the graphite nanoparticles, PCM, and glass cover cooling are used together, modification (D),, the water productivity of the solar still is increased by 73.8% over the conventional still.
5- The best performance is obtained for 0.5cm water depth for all solar stills.

## Method

**Experimental Setup:**

The solar stills and all components of the system were manufactured in the School of Energy and Power Engineering, Huazhong University of Science and Technology, Wuhan, China (Latitude 29°58°N and longitude 113°53°E).

Three solar stills with the same sizes were designed and manufactured to compare the performance of the solar desalination systems. A photograph and a schematic diagrams of the solar desalination setup are shown in Figs. 1 and 2 respectively. The system consists of a single basin solar still (conventional still), solar still with graphite nanofluid without and with glass film cooling (modified (A), (C) respectively), solar still with graphite nanofluid and PCM without and with glass film cooling (modified (B), (D) respectively), and film cooling water tank. Basin area of all stills are $0.25m^2$ (0.5m length × 0.5m width). The low-side wall height is 160mm and the high-side wall depth is 450mm. The stills are made of iron sheets (1.5mm thick). To increase the absorptivity of the solar still and hence increase the evaporation rate, a black paint is used to coat all basin surfaces from inside. To keep the heat loss as low as possible, all external surfaces and bottom are well insulated by fiberglass of 5cm thickness. The trough inside the basin still is used to collect the distillate output water into the external calibrated flasks through plastic pipes. The drain brine fluid is wasted outside the basin still through other pipes.

The basin was covered with a commercial clear glass sheet of 3.5mm thickness inclined at nearly 30° horizontally, which is the latitude of Wuhan, China. This tilt angle was selected to maximize the insulation received by the absorber and to minimize reflection losses. The whole experimental setup is kept in the south direction to receive maximum solar radiation throughout the year.

For modifications (A) and (C), the concentration of the graphite nanoparticles is 0.5%. For modifications (B) and (D), 20 pipes, coated with black paint from outside, are used and placed in the basin still. Each pipe has 49cm in length and 1.6cm in diameter and is used to contain the PCM (paraffin wax).

The cold water flowing over the glass (film cooling) was kept constant and uniform with the help of a constant head tank and a regulator. The cold water tank has the dimensions of 88×42×42cm. The glass cover, from inside direction, condenses the uprising evaporated water.

Due to the tilting of glass and gravity, the condensed water runs down through the small inclined triangular channel (trough) to be collected into the flasks. Basically, the brine and glass cover temperatures, ambient temperature, total solar radiation, wind velocity, and the amount of distillate are measured every 1 hour. The temperatures have been measured using calibrated copper constantan type thermocouples with range of (-50 to 180°C) with accuracy of (±1 °C) which are connected to a digital temperature indicator. While, solar meter range of (0-2000 W/m$^2$)with accuracy of ±10 W/m$^2$ is used to measure the solar intensity. The wind velocity is measured using the Vane type digital anemometer with range of (0-30 m/s) with accuracy of ±0.1m/s. Finally, a flask of 1.5 l capacity with accuracy of 5 ml is used to measure the productivity. The specifications of graphite nanoparticles and PCM are shown in Table 1. The flow rate of water used for cooling the glass cover is about 0.03 kg/s which is agreed with[25].

**Acknowledgements:**

N.Y. was sponsored by the National Natural Science Foundation of China (No. 51576076)12

**Author contributions:**

S.W.S, G.P. and L.W. Carried out the experimental measurements. N.Y. supervised the research. All authors analyzed the experimental data and edited the manuscript.

**Competing financial interests**: The authors declare no competing financial interests.

**Figures list**

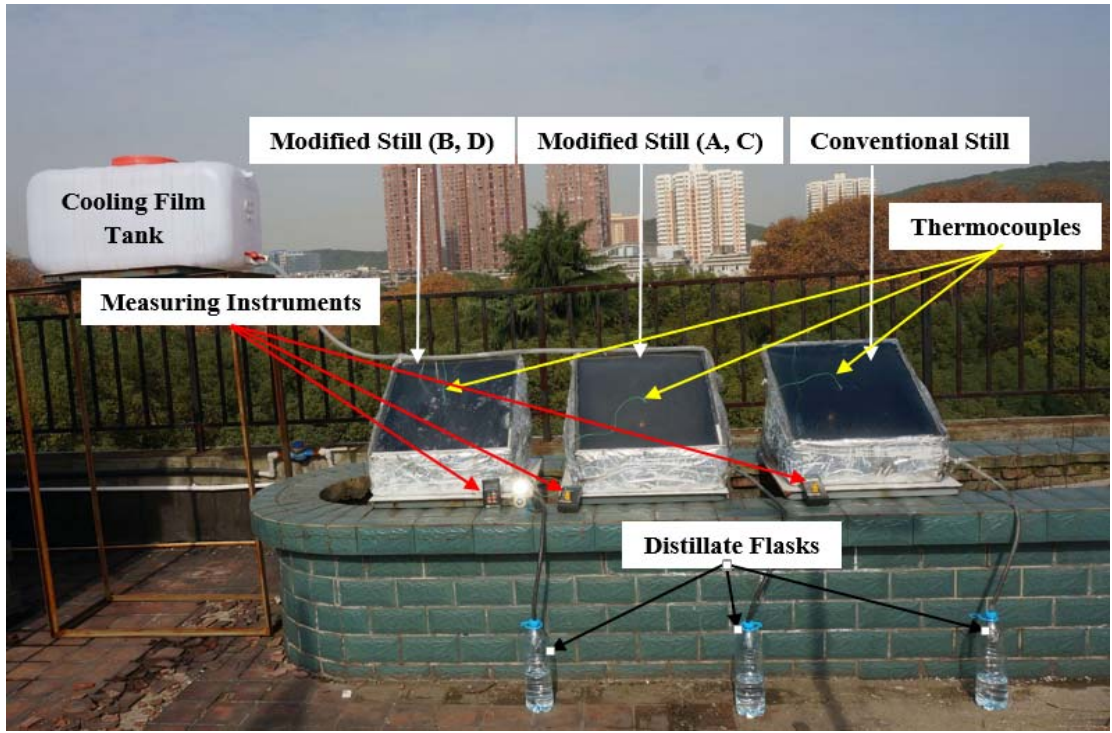

Fig.1. Photograph of the experimental setup.

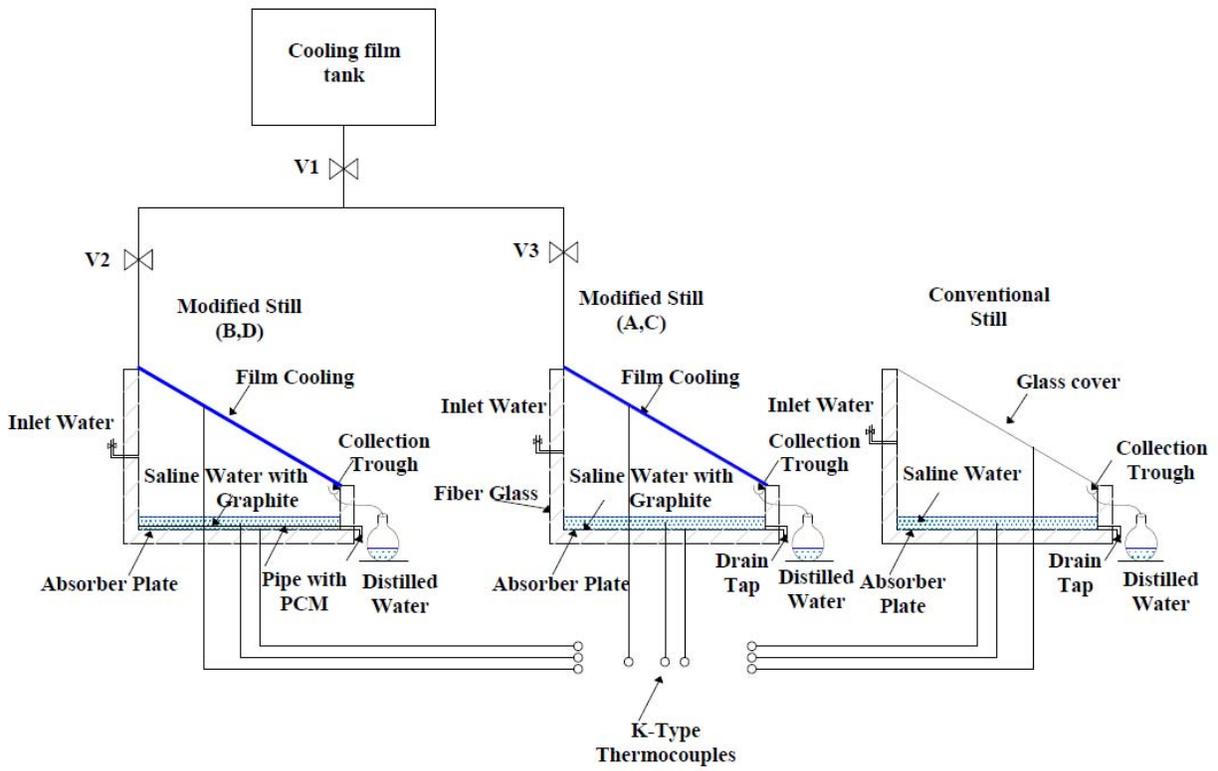

Fig.2. Schematic diagram of the experimental setup.

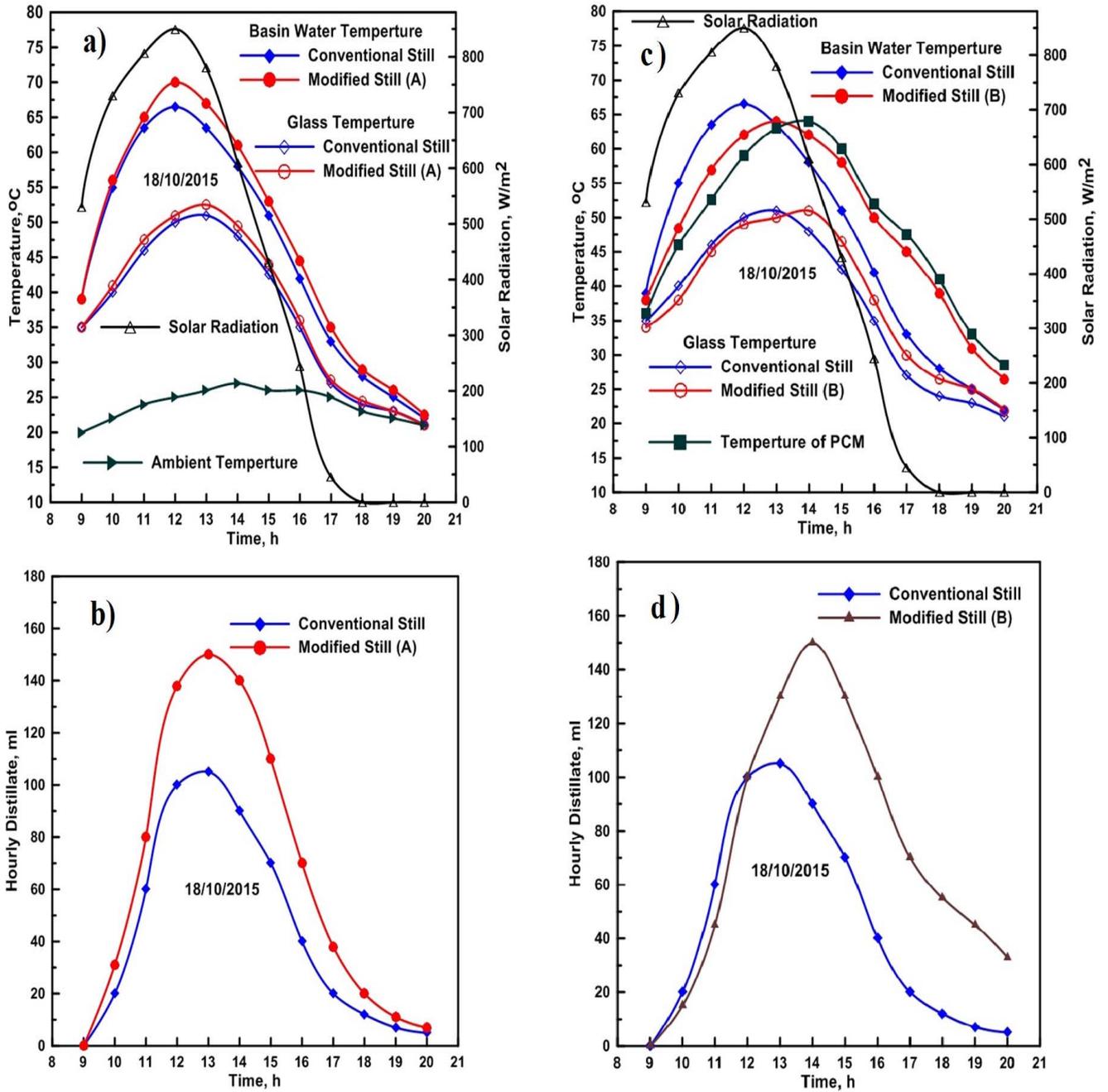

Fig. 3. Hourly variations of solar radiation, basin water , glass temperatures and productivity for the modified (A) and (B) with conventional stills. a), c). Solar radiation and temperatures, b), d). Fresh water productivity.

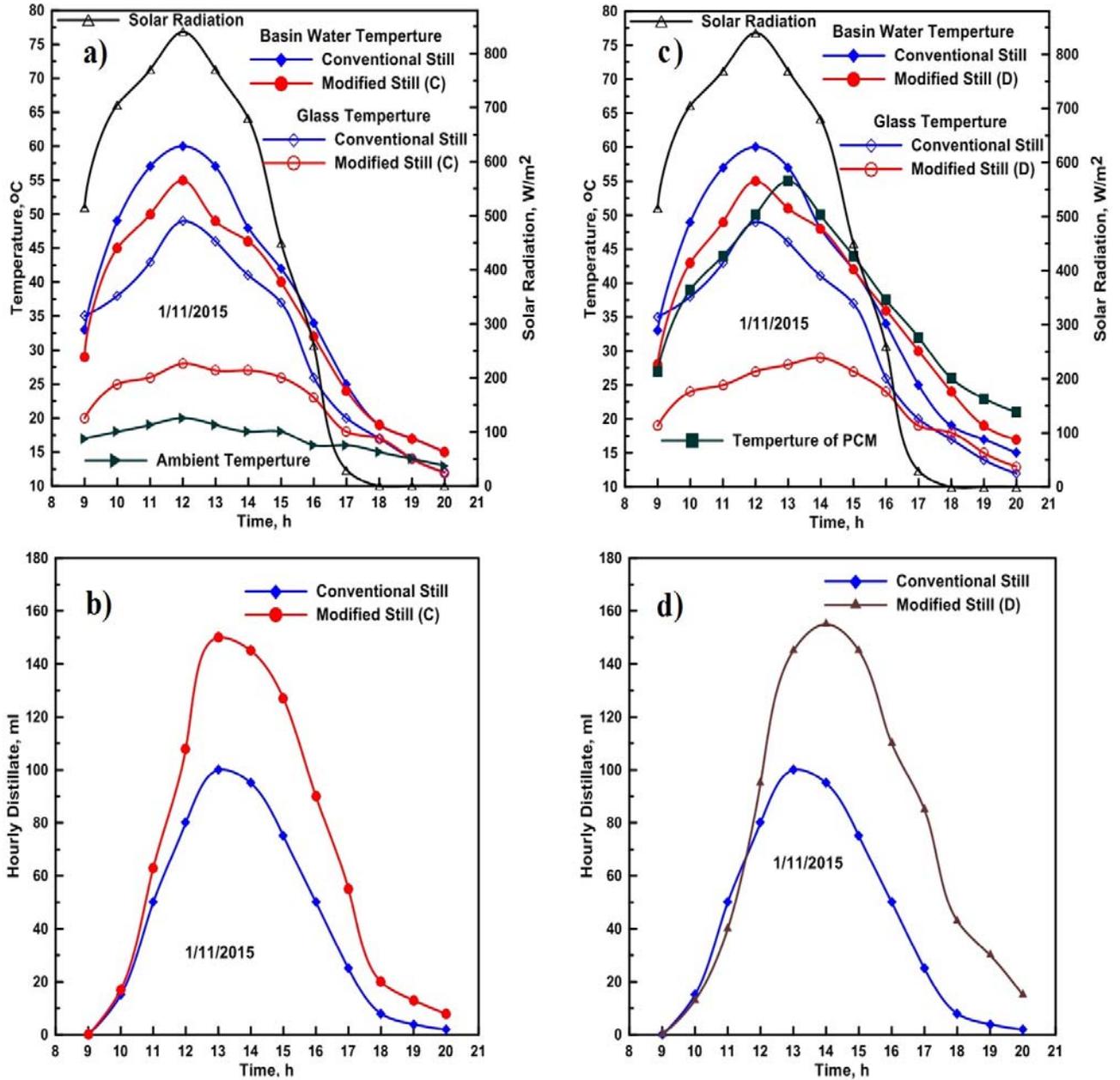

Fig. 4. Hourly variations of solar radiation, basin water, glass temperatures and productivity for the modified (c) and modified (d) conventional stills. a),c). Solar radiation and temperatures, b),d). Fresh water productivity.

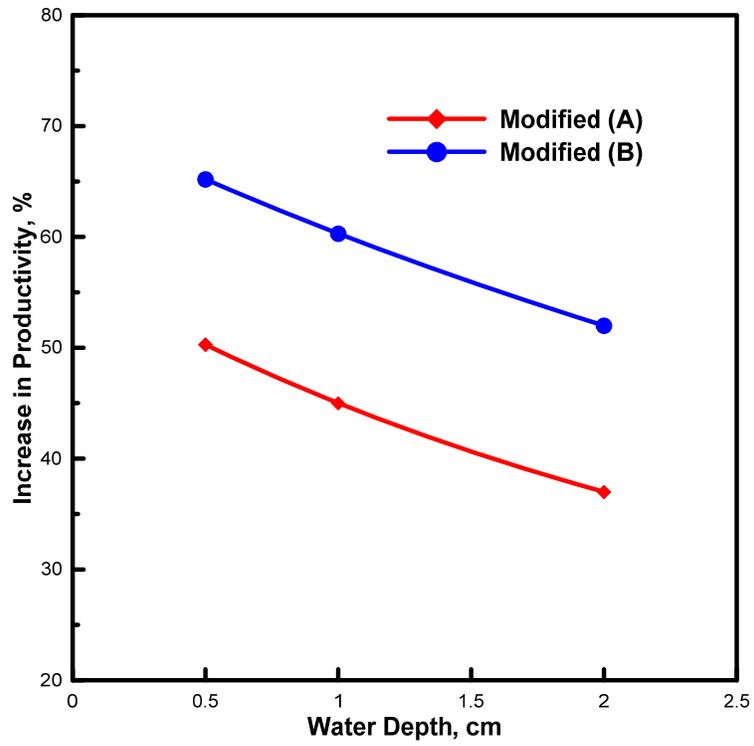

Fig. 5. Increase in productivity for the modified and the conventional stills for the two modes of testing.

Table list

Table 1 Specifications of graphite nanoparticles and PCM

| Property | Value |
|---|---|
| Thermal Conductivity, [W/(m.K)] | 129 |
| Density of nanoparticles, [g/cm$^3$] | 2 |
| Flake lateral size, [μm] | 1.2~1.3 |
| Concentration, [%] | 0.5 |
| Melting temperature of PCM, ºC | 48 |

Table 2. Comparison between present study and different research works about solar still with nanofluids

| References, location | Modification | Maximal. enhancement in productivity |
|---|---|---|
| Present study, Wuhan, China | Using the graphite nanoparticles [modification (A) phase change material (PCM) with graphite nanoparticles [modification (B)], graphite nanoparticles with glass film cooling [modification (C)]and PCM with graphite nanoparticles and glass film cooling [modification (D)] | 50.28%, modification(A). 65.00,%,modification (B). 56.15%, modification (C). 73.80%, modification (D). |
| Nijmeh et al[42], Amman, Jordan | Using potassium permanganate: ($KMnO_4$) and potassium dichromate ($K_2Cr_2O_7$) | 26%, with $KMnO_4$. 17%, with $K_2Cr_2O_7$. |
| Elango et al.[43], Tamil Nadu, India | Using Aluminum Oxide ($Al_2O_3$), Iron Oxide ($Fe_2O_3$), Zinc Oxide (ZnO) nanoparticles | 29.95%, Aluminum Oxide. 18.63%, Iron Oxide. 12.67%, Zinc Oxide. |
| Kabeel et al.[44], Kafrelsheikh, Egypt | Using the cuprous oxide and aluminum oxide nanoparticles with providing vacuum | 133.64% cuprous oxide with vacuum. 125.0% aluminum oxide with vacuum. |
| Kabeel et al.[45], Kafrelsheikh, Egypt | Using aluminum-oxide nanoparticles and external condenser | 116% aluminum-oxide with vacuum. |
| Sahota and Tiwari[46] New Delhi, India | Using aluminum-oxide nanoparticles | 12.2% aluminum oxide |